\documentclass{emulateapj}
\usepackage{graphicx}
\usepackage{amsmath}
\usepackage{natbib}
\usepackage[usenames]{color}
\usepackage{float}
\usepackage{subfigure}
\usepackage{hyperref}
\usepackage{color}

\newcommand{\sauron}{\texttt{SAURON}}
\newcommand{\atlas}{\texttt{ATLAS$^{\rm 3D}$}}
\newcommand{\califa}{\texttt{CALIFA}}

\newcommand{\galaxy}{\texttt{GALAXY}}

\newcommand{\shump}{\hbox{$\sigma$-humps}}
\newcommand{\shollow}{\hbox{$\sigma$-hollows}}
\newcommand{\sLOS}{\hbox{$\sigma$}}
\newcommand{\szhump}{\hbox{$\sigma_z$-humps}}

\newcommand{\shuho}{\hbox{$\sigma$-humps+hollows}}
\newcommand{\szhuho}{\hbox{$\sigma_z$-humps+hollows}}
\newcommand{\refsec}[1]{Section~\ref{#1}}
\newcommand{\reffig}[1]{Fig.~\ref{#1}}
\newcommand{\refeq}[1]{Eqn.~\ref{#1}}
\newcommand{\reftab}[1]{Table~\ref{#1}}

\begin{document}
\title{Orthogonal vertical velocity dispersion distributions produced by bars}
\author{
Min Du\altaffilmark{1, 2}, 
Juntai Shen\altaffilmark{1, 2},
Victor P. Debattista\altaffilmark{2, 3},
Adriana de Lorenzo-C{\'a}ceres\altaffilmark{4}
}

\altaffiltext{1}{Key Laboratory of Research in Galaxies and Cosmology, Shanghai Astronomical Observatory, Chinese Academy of Sciences, 80 Nandan Road, Shanghai 200030, China}
\altaffiltext{2}{Correspondence should be addressed to dumin@shao.ac.cn; jshen@shao.ac.cn; VPDebattista@uclan.ac.uk}
\altaffiltext{3}{Jeremiah Horrocks Institute, University of Central Lancashire, Preston, PR1 2HE, UK}
\altaffiltext{4}{Instituto de Astronomia, Universidad Nacional Aut$\acute{o}$noma de M$\acute{e}$xico, A. P. 70-264, 04510 Mexico City, Mexico}

\begin{abstract}
In barred galaxies, the contours of stellar velocity dispersions (\sLOS) are generally expected to be oval and 
aligned with the  
orientation of bars. However, many double-barred (S2B) galaxies exhibit distinct \sLOS\ peaks on the minor axis of the inner bar, which we 
termed ``\shump,'' while two local \sLOS\ minima are present close to the ends of inner bars, i.e., ``\shollow.'' 
Analysis of numerical simulations shows that \szhump\ or hollows should play an important role in 
generating the observed \shuho\ in low-inclination galaxies. In order to systematically investigate the properties of $\sigma_z$ in barred 
galaxies, we apply the vertical Jeans equation to a group of well-designed three-dimensional bar+disk(+bulge) models. 
A vertically thin bar can lower $\sigma_z$ along the bar and enhance it perpendicular to the bar, thus generating 
\szhuho. Such a result suggests that \szhuho\ can be generated by the purely dynamical response of stars in the presence of a sufficiently 
massive, vertically thin bar, even without an outer bar. Using self-consistent $N$-body simulations, we verify the existence 
of vertically thin bars in the nuclear-barred and S2B models that generate prominent \shuho. 
Thus, the ubiquitous presence of \shuho\ in S2Bs implies that inner bars are vertically thin. The addition of a bulge makes
the \szhump\ more ambiguous and thus tends to somewhat hide the \szhuho. We show that $\sigma_z$ may 
be used as a kinematic diagnostic of stellar components that have different thickness, providing a direct  
perspective on the morphology and thickness of nearly face-on bars and bulges with integral field unit spectroscopy. 
\end{abstract}

\keywords{galaxies: kinematics and dynamics --- galaxies: structure --- galaxies: stellar content --- galaxies: bulges }

\section{Introduction}
\label{introduction}

Near-infrared imaging surveys have shown that bars are ubiquitous stellar structures; in the local universe about 
two-thirds of disk galaxies host elongated stellar bars \citep{Eskridge00, Whyte02, Laurikainen04a, Marinova&Jogee07, Menendez-Delmestre07}. 
The fraction is $0.25-0.3$ if only strong bars are counted \citep[e.g.][]{Nilson73, deVaucouleurs91}. 
From $N$-body simulations it is well known that bars can spontaneously form in galactic disks if the disk dynamical temperature 
(Toomre-$Q$) is not too high \citep[e.g.][]{Miller70, Hohl71, ost_pee_73, sel_80, sel_81, Athanassoula&Sellwood86}. Once formed, bars are 
expected to be long-lived and difficult to destroy \citep{shen04, Debattista06, villa-vargas10, athanassoula13}, which is supported by 
the fact that bars are typically composed of old stars \citep{gadotti06, sanchez11}. Observations of intermediate-redshift galaxies have
revealed that the fraction of bars increases from $\sim20\%$ at $z\sim0.84$ to $\sim65\%$ in the local universe \citep{sheth08}. 
As the frequency of violent interactions between galaxies decreases, the evolution of galaxies is driven mainly by internal processes, 
so-called secular evolution. Galactic bars are the most important driver of the secular evolution of disk galaxies 
\citep[see the reviews of][]{Kormendy04, Kormendy13}. By transferring angular momenta to the outer disk and dark matter halo, bars may 
grow longer and stronger, but rotate slower \citep{Debattista&Sellwood98, Debattista&Sellwood00, ath_03}. Bars can drive the transport 
and accumulation of gas toward galactic central regions, thus triggering nuclear starbursts and, possibly, fueling active galactic nuclei 
(AGN) \citep[e.g.][]{Shlosman89, Shlosman90, Buta&Combes96, Bournaud&Combes02, Maciejewski04a, Maciejewski04b, Garcia-Burillo05, Hopkins10a, Kim12a, Emsellem15, LiZhi15}. 
Numerical simulations also suggest that bar formation can trigger the vertical buckling instability, leading to boxy/peanut (B/P) 
bulges \citep{rah_etal_91, Merritt&Sellwood94}. 

Being composed primarily of old stars, bars can be traced well in infrared bands where the dust extinction is much weaker than that 
in visible bands. The morphology of bars has been systematically investigated through ellipse fitting and Fourier decomposition of 
infrared images \citep[e.g.][]{Chapelon99, Knapen00, Laine02, Laurikainen&Salo02, Laurikainen02, Erwin05}. 
Early dynamical studies of bars used long-slit spectroscopy of stars and ionized gas 
\citep[e.g.][]{Kuijken&Merrifield95, Bureau&Freeman99, VegaBeltran01}. In the past decade, the development of integral field unit 
(IFU) spectroscopy has made it possible to study the 2D kinematics of nearby galaxies. The kinematics of disks and bars have been quantified 
in several IFU surveys, e.g. \sauron\ \citep{deZeeuw_etal_02}, \atlas\ \citep{cap_etal_11}, \califa\ \citep{san_etal_12}, DiskMass 
\citep{Bershady10a}, and MaNGA \citep{Bundy15} \citep[see review by][]{cap_16}. However, knowledge of the kinematic properties of 
bars is still incomplete. 
In early-type barred galaxies the central kinematic major axis is misaligned from the line of nodes (LON) by around $\sim 5^\circ$
\citep{Cappellari07, Krajnovic11}. This is probably because the elongated streaming motions in bars distort the velocity fields, as 
shown in numerical studies \citep{Miller&Smith79, Vauterin&Dejonghe97, Bureau&Athanassoula05}. According to $N$-body 
simulations, the kinematic misalignment is not prominent in bars of early-type galaxies, possibly because large random motions dominate
\citep{Du16}. Generally, face-on or moderately inclined bars are expected to generate oval velocity dispersion contours aligned with 
the bar \citep{Debattista05, ian_ath_15, Du16}. In this paper we simply refer to the line-of-sight (LOS) velocity dispersion 
$\sigma_{\rm LOS}$ as \sLOS. 

The most surprising \sLOS\ features are the \shump\ and hollows found in double-barred (S2B) galaxies. 
Using \sauron\ IFU spectroscopy, \citet{deLorenzoCaceres08} found that the \sLOS\ maps of S2Bs reveal two local minima at the ends of 
inner bars, which they termed ``\shollow.'' \citet{Du16} showed that such \shollow\ can be reproduced in the 
self-consistent simulations of S2Bs, which match well the \sLOS\ features in the S2Bs in the \atlas\ and \sauron\ surveys. The 
S2B simulations exhibit double-peaked \sLOS\ enhancements along the minor axis of inner bars as well, termed 
``\shump,'' which are also seen in the observations. 
Optical and near-infrared observations have revealed that multibar structures are quite common in the local universe; almost one-third of 
early-type barred galaxies host S2B structures \citep{Erwin&Sparke02, Laine02, Erwin04}. Many observations have shown that in 
S2Bs small-scale inner bars are dynamically decoupled from their large-scale outer counterparts
\citep{Buta&Crocker93,Friedli&Martinet93, Corsini03}. Inner bars are generally expected to rotate faster than outer ones.
The physical origin of \shuho\ is still unclear. \citet{Du16} reported that \shuho\ often accompanied nuclear bars in single-barred models. 
Therefore, \shuho\ are not unique features of S2Bs, and cannot arise from the interaction of two bars. \citet{deLorenzoCaceres08} proposed that 
\shollow\ are simply caused by the contrast of a dynamically cold inner bar embedded in a relatively hotter bulge. In \citet{Du16} 
we analyzed the difference in intrinsic kinematics between the model with \shuho\ and that without \shuho. Their only difference 
is the double-peaked vertical velocity dispersion ($\sigma_z$) enhancements perpendicular to the inner bar, i.e., \szhump, which must play 
an important role in generating \shuho.

In this paper, we investigate the $\sigma_z$ properties in a family of analytical models of barred galaxies, i.e., bar+disk systems, which 
are introduced in \refsec{IC}. The analytical results are presented in \refsec{results}, where we successfully explain the physical origin 
of \szhuho\ from a purely dynamical point of view. In \refsec{bulge}, we test the effect of a massive bulge component on $\sigma_z$ 
features. In \refsec{simulation}, using the self-consistent nuclear-barred and large-scale single-barred simulations, we verify the 
analytical results. We further demonstrate that $\sigma_z$ can be used as a kinematic diagnostic of the 
relative thickness of different stellar components. Finally, we summarize the conclusions of this work 
in \refsec{summary}.

\section{Method}
\label{IC}

\subsection{Vertical Jeans equation}
\label{Jeans}

Galactic disks are equilibrium systems whose stellar kinematics must satisfy the Jeans equations \citep[][Equation (4.208)]{Binney&Tremaine08}  
which were first applied to stellar systems by \citet{jea_22}. 
In the coordinate system rotating with bar pattern speed $\bf{\Omega}_p$ about the $z$-axis, the fictitious forces must be 
considered. The Coriolis and centrifugal forces on one mass unit are $-2\bf{\Omega}_p \times \bf{v}$ and $-\Omega_p^2 \bf{R}$, 
respectively. In the Cartesian coordinate system, the streaming motion vector $\bf{v}$ is $(\overline{v_x}, \overline{v_y}, \overline{v_z})$. 
Thus, the Jeans equations are written as
\begin{equation}
\begin{aligned}
        \centering
        \frac{\partial (\rho_\star \overline{v_{x}^2})}{\partial x} + \frac{\partial (\rho_\star \overline{v_x v_y})}{\partial y} + \frac{\partial (\rho_\star \overline{v_x v_z})}{\partial z} = & \\ - \rho_\star \frac{\partial \Phi}{\partial x} - x \rho_\star \Omega_p^2 + 2\rho_\star \Omega_p\overline{v_y}  \\ 
        \frac{\partial (\rho_\star \overline{v_x v_y})}{\partial x} + \frac{\partial (\rho_\star \overline{v_y^2})}{\partial y} + \frac{\partial (\rho_\star \overline{v_y v_z})}{\partial z} = & \\ - \rho_\star \frac{\partial \Phi}{\partial y} - y \rho_\star \Omega_p^2 - 2\rho_\star \Omega_p\overline{v_x}  \\ 
        \frac{\partial (\rho_\star \overline{v_x v_z})}{\partial x} + \frac{\partial (\rho_\star \overline{v_y v_z})}{\partial y} + \frac{\partial (\rho_\star \overline{v_z^2})}{\partial z} = & - \rho_\star \frac{\partial \Phi}{\partial z}, 
\end{aligned}
        \label{eq:jeans}
\end{equation}
where $\Phi$ is the total potential, including the contributions from the stellar component and dark matter halo, and $\rho_\star$ is the 
volume density of the stellar component. 

For disk galaxies, the Jeans equations generally cannot be uniquely solved in the 
disk plane (the $x-y$ plane in Cartesian coordinates) without assumptions. In this paper we are only concerned with the vertical 
Jeans equation, i.e., the $z$-direction, which can be written as 
\begin{equation}
        \centering
        \frac{\partial (\rho_\star \sigma_{xz}^2)}{\partial x} + \frac{\partial (\rho_\star \sigma_{yz}^2)}{\partial y} + \frac{\partial (\rho_\star \sigma_{z}^2)}{\partial z} = - \rho_\star \frac{\partial \Phi}{\partial z} = \rho_\star F_z, 
        \label{eq:jeansz}
\end{equation}
where $\sigma_{iz}^2 = \overline{v_iv_z} - \bar{v_i}\bar{v_z} = \overline{v_iv_z}$; $i=x, y, z$, assuming that $\overline{v_z}$ is 
always zero. Generally, $\overline{v_z}=0$ is expected to be satisfied except when the disk is undergoing significant 
buckling or bending motions. 
$\sigma_{zz}^2$ is written as $\sigma_{z}^2$ for short. $F_z$ is the vertical gravitational force. With the 
boundary condition $\rho_\star=0$ as $z \rightarrow \infty$, the integral from $z$ to $\infty$ gives 
\begin{equation}
\begin{aligned}
        \rho_\star(z) \sigma_{z}^2(z) = & -\int_{z}^{\infty} \rho_\star(z') F_{z}(z') \mathrm{d}z' \\ 
         & + \int_{z}^{\infty} \left[\frac{\partial (\rho_\star \sigma_{xz})}{\partial x} + \frac{\partial (\rho_\star \sigma_{yz})}{\partial y}\right]\mathrm{d}z'.
      \label{eq:int}
\end{aligned}
\end{equation}
Corresponding to the anisotropic pressure forces, the second-order velocity moments, $\sigma_{xz}$ and $\sigma_{yz}$, are omitted in the
following analyses; thus,  
\begin{equation}
        \centering
        \rho_\star(z) \sigma_z^2(z) \approx -\int_{z}^{\infty} \rho_\star(z') F_{z}(z') \mathrm{d}z'.
        \label{eq:sJeans}
\end{equation}
The legitimacy of this assumption in barred galaxies will be discussed in \refsec{aniso}. 
The integral of the velocity dispersion in the face-on view is obtained by
\begin{equation}
        \begin{aligned}
        \left<\sigma_z^2\right> &= \frac{\int_{-\infty}^{+\infty} \rho_\star \sigma_z^2 \mathrm{d}z}{\int_{-\infty}^{+\infty} \rho_\star \mathrm{d}z}
                      = \frac{\int_{-\infty}^{+\infty} \rho_\star \sigma_z^2 \mathrm{d}z}{\Sigma} \\
                     &\approx \frac{-\int_{-\infty}^{+\infty} \int_{z}^{\infty} \rho_\star(z') F_{z}(z') \mathrm{d}z'\mathrm{d}z}{\Sigma}. 
        \end{aligned}
\label{eq:zint}
\end{equation}
Therefore, the vertical dynamics give a simple relation between the density distribution of stars and the vertical velocity 
dispersion $\sigma_z$. Assuming a reasonable vertical density distribution, the axisymmetric form of this relation has been used as 
a kinematic estimator of the stellar disk mass in the DiskMass IFU survey \citep{Bershady10a, Bershady11, Martinsson13a, Martinsson13b, Angus15}. 
In the DiskMass survey, the influence of bars is generally ignored by selecting a sample of unbarred or small/weakly barred galaxies. 

\subsection{Equilibrium bar+disk models}

\begin{deluxetable}{lcccc}
\tablecaption{Settings of the bar+disk systems}
\tablenum{1}
\tablehead{\colhead{Name} & \colhead{$h_z$} & \colhead{$M_\mathrm{B}$} & \colhead{Ferrers $n$} & \colhead{$a/b/c$}} 
      \startdata
        E0 &  0.3   & 0    & ...         & ... \\
        E1 &  0.3   & 0.01 & 1.0         & 1.0/0.4/0.1 \\    
        E2 &  0.3   & 0.03 & 1.0         & 1.0/0.4/0.1 \\    
        E3 &  0.3   & 0.05 & 1.0         & 1.0/0.4/0.1 \\ \hline    
        L1 &  0.1   & 0.05 & 1.0         & 1.0/0.4/0.1 \\  
        L2 &  0.2   & 0.05 & 1.0         & 1.0/0.4/0.1 \\ \hline
        E4 &  0.3   & 0.05 & 1.0         & 1.0/0.2/0.1 \\  
        E5 &  0.3   & 0.05 & 1.0         & 1.0/0.6/0.1 \\ \hline
        E6 &  0.3   & 0.05 & 1.0         & 1.0/0.4/0.3 \\  
        E7 &  0.3   & 0.05 & 1.0         & 1.0/0.4/0.5 \\ \hline
        E8 &  0.3   & 0.05 & 1.0         & 2.0/0.4/0.1 \\  
        E9 &  0.3   & 0.20 & 1.0         & 2.0/0.4/0.1 \\
        E10&  0.3   & 0.20 & 1.0         & 2.0/0.4/0.3 
      \enddata
\tablecomments{From left to right: model name, disk scale height $h_z$, bar mass $M_\mathrm{B}$, Ferrers $n$, and axial ratio of bars $a/b/c$.}
\label{tab:modelIC}
\end{deluxetable}

\begin{figure*}[htp]
\centering
       \subfigure{\includegraphics[width=0.95\textwidth]{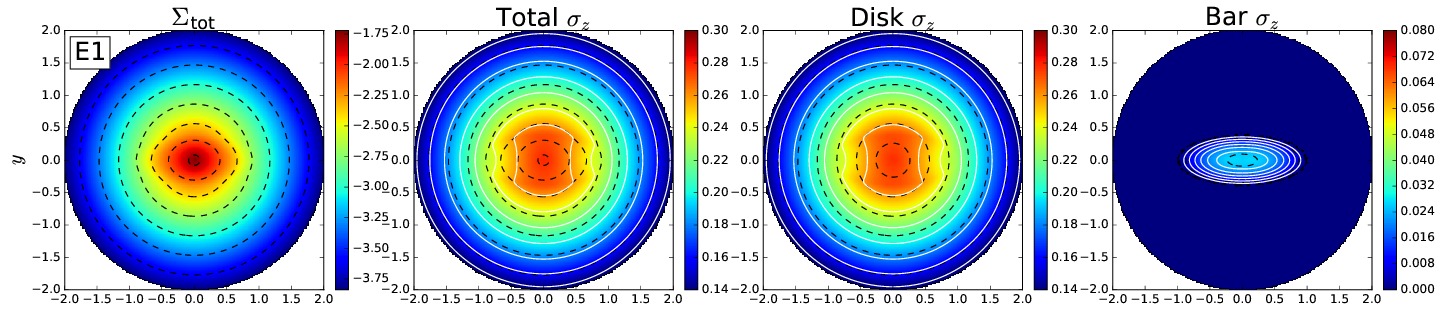}}
       \subfigure{\includegraphics[width=0.95\textwidth]{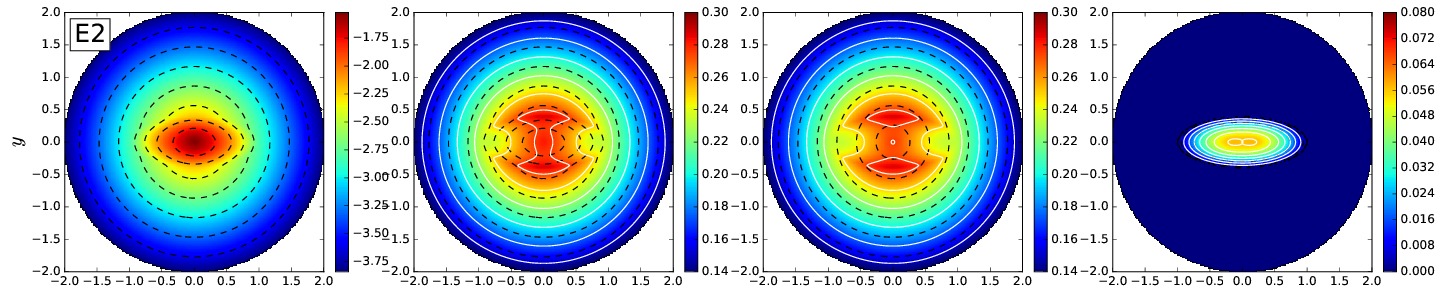}}
       \subfigure{\includegraphics[width=0.95\textwidth]{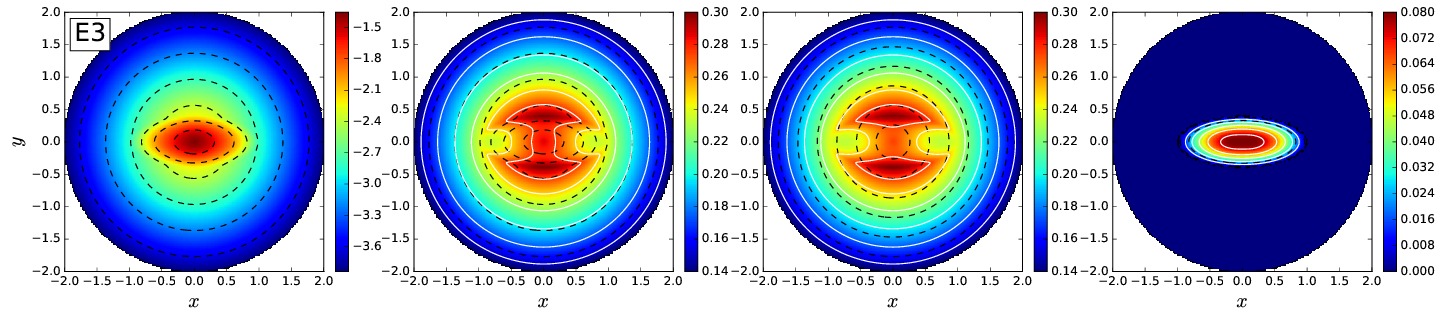}}
       \caption{\label{fig:BxDs}Total surface density $\Sigma_{\rm tot}$ and $\sigma_z$ maps of the models E1-E3 in 
               \reftab{tab:modelIC}, showing the variation of $\sigma_z$ fields with increasing bar mass. From top to bottom, 
               the bar mass is set to $0.01M_\mathrm{D}$ (E1), $0.03M_\mathrm{D}$ (E2), and $0.05M_\mathrm{D}$ (E3), respectively. 
               From left to right: $\Sigma_{\rm tot}$, total $\sigma_z$, disk $\sigma_z$, and bar $\sigma_z$. 
               The isodensity contours are equally separated in logarithmic space, and $\sigma_z$ contours of each map are overlaid using 
               black dashed and white solid curves, respectively.}
\end{figure*}

\begin{figure}[htp]
\centering
       \subfigure{\includegraphics[width=0.45\textwidth]{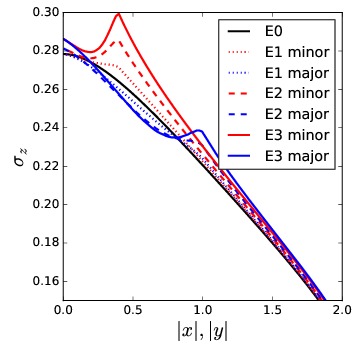}}
       \caption{\label{fig:1DBxD}1D $\sigma_z$ profiles along the minor (red) and major (blue) axes of the bars in the models E0-E3, varying 
            the bar mass $M_\mathrm{B}$. The black solid profile shows the $\sigma_z$ profile of the pure-disk model E0.} 
\end{figure}

A $\sigma_z$ map for any arbitrary density distribution can be numerically computed using \refeq{eq:zint}. In order to study 
the $\sigma_z$ features in barred galaxies, we use a family of bar+disk models that are embedded in a dark matter 
potential. In cylindrical coordinates, the disk density we use is given by a simple double-exponential distribution 
\begin{equation}
        \centering
	\rho_\mathrm{D}(R, z) = \frac{\Sigma_0}{2 h_z} \mathrm{exp}(-\frac{R}{h_R}-\frac{z}{h_z}),
\end{equation}
where $\Sigma_0, h_R$, and $h_z$ are the central surface density, scale length, and scale height, respectively, of the disk 
component. Thus the disk mass $M_\mathrm{D}$ is $2\pi \Sigma_0 h_R^2$. To simplify the following analyses, we use the same unit system 
as in \citet{Du15}, i.e., $M_\mathrm{D}=G=h_R=V_0=1$. 

The bar component is modeled as a triaxial Ferrers ellipsoid \citep{Ferrers1877}
\begin{equation}
\centering
	\rho_\mathrm{B}(x, y, z)=\left\{
	\begin{aligned}
		 & \rho_{\rm B0} (1 - m^2)^n & m \le 1  \\
		 & 0 & m > 1, \\
\end{aligned}
\right.
\end{equation}
where $m^2=x^2/a^2 + y^2/b^2 + z^2/c^2$ in Cartesian coordinates. The bar is aligned along the $x$-axis; the values of $a, b$, 
and $c$ determine the semimajor axis, semiminor axis, and thickness, respectively, of the bar. The axial ratio $b/a$ corresponds 
to the ellipticity ($\epsilon$) of the bar. The Ferrers $n$ parameter determines how fast the density decreases outward. 
Photometric observations show that the typical density profiles of bars are shallow (flat) and clearly 
truncated in early-type galaxies, while in late-type galaxies they tend to decrease outward, following a more exponential profile 
\citep{Elmegreen&Elmegreen85, Chapelon99, Laine02, Laurikainen&Salo02, Laurikainen02, Kim16}. 
The central density of the bar, $\rho_{\rm B0}$, is numerically calculated for a given total mass of the bar, $M_{\rm B}$. We use the 
same dark matter potential as in the $N$-body simulations of \citet{Du15},
\begin{equation}
	\centering
	\Phi_\mathrm{DM} = \frac{1}{2} V_h^2 \ln (r^2 + r_h^2),
	\label{DM}
\end{equation}
where $V_h=0.6$ and $r_h = 15$.

In such bar+disk systems, there are six free parameters in total: the disk scale height $h_z$, bar mass $M_\mathrm{B}$, and four structural 
parameters of the bar $a, b, c$, and Ferrers $n$. The Ferrers $n$ is fixed at 1.0 to generate a shallow bar model.
Given a set of parameters, we use an $N_x \times N_y \times N_z=201\times201\times401$ 
grid to calculate the gravitational potential of the total system with a parallel 3D Poisson solver PSPFFT \citep{Budiardja11}. 
The model is located at the geometric center of the grid, and the spatial resolution is constant at 0.01 along the $z$-direction as 
$z\in[-2.0, 2.0]$. Including the dark matter potential, we numerically compute $\sigma_z$ using \refeq{eq:zint}. 

Systematic studies of near-infrared images of barred galaxies have shown that the semimajor axis of bars 
varies up to $2.5h_R$, with the mean values $\sim1.3h_R$ and $\sim0.6h_R$ for early-type and 
late-type galaxies, respectively \citep{Erwin05, Menendez-Delmestre07, Diaz-Garcia16}. The scale length of disk galaxies seems to be 
independent of their Hubble type \citep{deJ_96, gra_deB_01, fat_etal_10}.
However, optical and near-infrared observations of edge-on galaxies find a decreasing trend of 
the scale height $h_z$ from early-type to late-type galaxies \citep{deGrijs98, Schwarzkopf&Dettmar00, Bizyaev14, Mosenkov15}. 

\reftab{tab:modelIC} shows the set of parameters we choose to study the $\sigma_z$ features of barred galaxies. 
According to the empirical relation of \citet{deGrijs98}, the ratio $h_R/h_z$ varies from $1-5$ (the median value $\sim4$) in 
early-type spirals ($T\le1$) to $5-12$ (the median value $\sim9$) in very late-type 
spirals ($T\ge5$). Thus, we vary $h_z$ from 0.3 to 0.1 in our analytical models. In the case of $h_z=0.3$, we denote such early-type-like 
(``E'') double-exponential models as ``E*.'' The late-type-like double-exponential models are denoted as ``L*'' 
in the cases of $h_z=0.1$ and $0.2$. The model E0 is a purely axisymmetric disk without a bar, i.e., $M_\mathrm{B}=0$. A typical bar is used 
in most models by setting $a=h_R=1.0$ and $b/a=0.4$. In the models E1-E7 and L1-L2 the bar length ($a=1.0$) is half of that in E8-E10 
($a=2.0$). We truncate the disk at twice the bar length in order to reduce the calculation time, which allows us to obtain a high enough
spatial resolution in the $x-y$ plane (0.02 in E0-E7 and L1-E2; 0.04 in E8-E10). We have confirmed that using a larger truncation 
radius makes negligible difference in the $\sigma_z$ map. The dark matter potential also has a tiny effect on $\sigma_z$.

To better understand the $\sigma_z$ features, we further decompose the contributions of the bar and the disk components as follows:
\begin{equation}
	\centering
	\left<\sigma_z^2\right> = \frac{\Sigma_\mathrm{D}}{\Sigma_{\rm tot}}\left<\sigma_z^2\right>_\mathrm{D} + \frac{\Sigma_\mathrm{B}}{\Sigma_{\rm tot}}\left<\sigma_z^2\right>_\mathrm{B}, 
	\label{eq:sigmaz2}
\end{equation}
where $\Sigma_{\rm D}$, $\Sigma_{\rm B}$, and $\Sigma_{\rm tot}$ are the surface densities of the disk, bar, and total stellar system, 
respectively, and $\left<\sigma_z\right>_\mathrm{D}$ and $\left<\sigma_z\right>_\mathrm{B}$ are the intrinsic vertical velocity dispersions of 
the disk and the bar, respectively. According to \refeq{eq:zint}, $\left<\sigma_z^2\right>_\mathrm{D}$ and $\left<\sigma_z^2\right>_\mathrm{B}$ 
can be calculated as 
\begin{equation}
        \begin{aligned}
        \left<\sigma_z^2\right>_\mathrm{D}  
                     &= \frac{-\int_{-\infty}^{+\infty} \int_{z}^{\infty} \rho_{\rm D} F_{z} \mathrm{d}z'\mathrm{d}z}{\Sigma_{\rm D}} \\
        \left<\sigma_z^2\right>_\mathrm{B}  
                     &= \frac{-\int_{-\infty}^{+\infty} \int_{z}^{\infty} \rho_{\rm B} F_{z} \mathrm{d}z'\mathrm{d}z}{\Sigma_{\rm B}}, \\
        \end{aligned}
\end{equation}
where $F_z$ is the total vertical force. In following analyses, we weight the disk and the bar $\sigma_z$ by the surface density, i.e., the disk 
$\sigma_z=\sqrt{\Sigma_\mathrm{D} \left<\sigma_z^2\right>_\mathrm{D}/\Sigma_{\rm tot}}$ and the bar 
$\sigma_z =\sqrt{\Sigma_\mathrm{B} \left<\sigma_z^2\right>_\mathrm{B}/\Sigma_{\rm tot}}$. Thus, the disk and the bar $\sigma_z$ are
their respective contributions to the total $\sigma_z$. The bulge $\sigma_z$ will be defined similarly when a bulge is added. As shown 
in \reftab{tab:modelIC}, we explore the effect 
of bar mass $M_\mathrm{B}$ (E0-E3), ellipticity $a/b$ (E3-E5), thickness $c$ (E3, E6-E7), and semimajor axis $a$ (E3, E8-E10) 
of the Ferrers bar. E3 exhibits prominent \shuho. Models L1-L2 show the effect of disk scale height $h_z$. 

\begin{figure*}[htp]
\centering
       \subfigure{\includegraphics[width=0.95\textwidth]{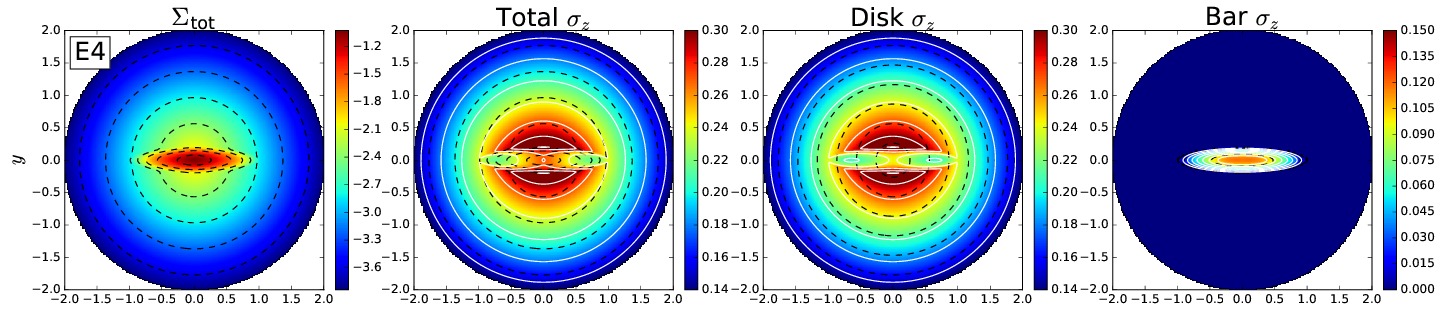}}
       \subfigure{\includegraphics[width=0.95\textwidth]{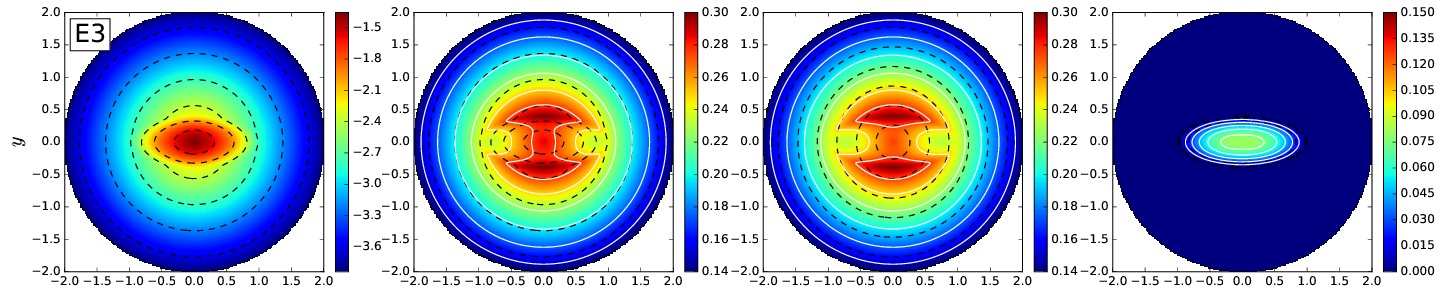}}
       \subfigure{\includegraphics[width=0.95\textwidth]{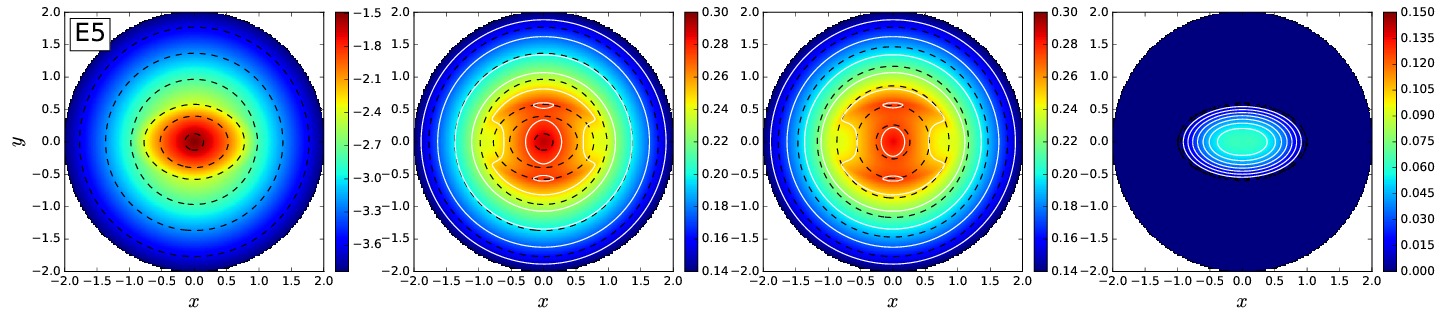}}
       \caption{\label{fig:a2}Models E4, E3, and E5 in \reftab{tab:modelIC}, showing the variation of $\sigma_z$ fields when varying the 
minor-to-major axial ratio $b/a$ to 0.2 (E4), 0.4 (E3), and 0.6 (E5).}
\end{figure*}
\begin{figure}[htp]
\centering
       \subfigure{\includegraphics[width=0.45\textwidth]{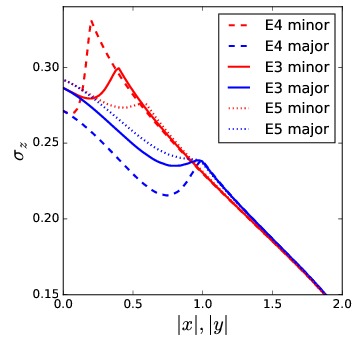}}
       \caption{\label{fig:1Da2}1D $\sigma_z$ profiles along the minor (red) and major (blue) axes of the bars in the models E4, E3 and E5, 
            varying the minor-to-major axial ratio $b/a$ of the bar.} 
\end{figure}



\section{Analytical Results: The Orthogonal $\sigma_z$ Features in Bar+Disk Systems}
\label{results}

\begin{figure*}[htp]
\centering
       \subfigure{\includegraphics[width=0.95\textwidth]{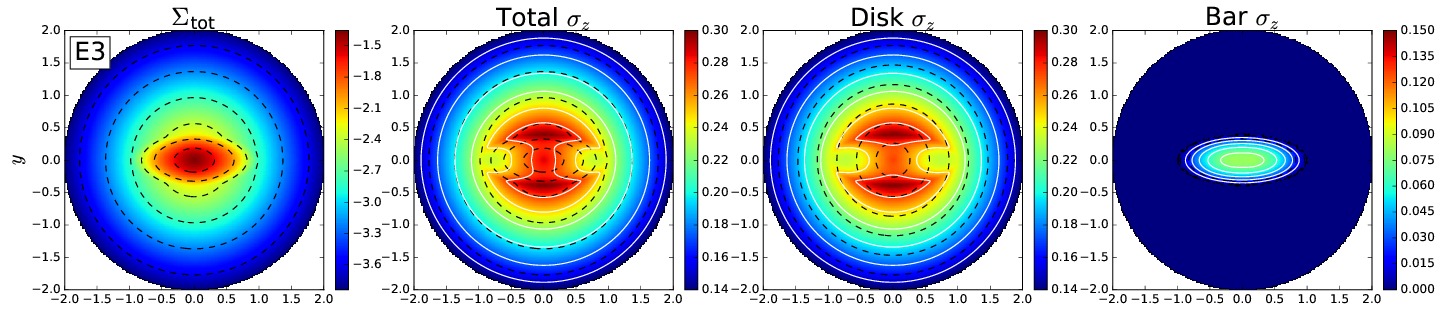}}
       \subfigure{\includegraphics[width=0.95\textwidth]{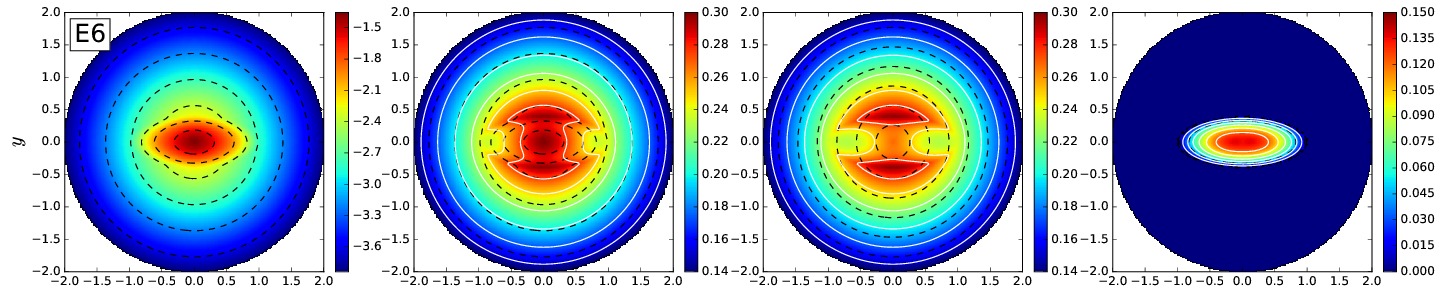}}
       \subfigure{\includegraphics[width=0.95\textwidth]{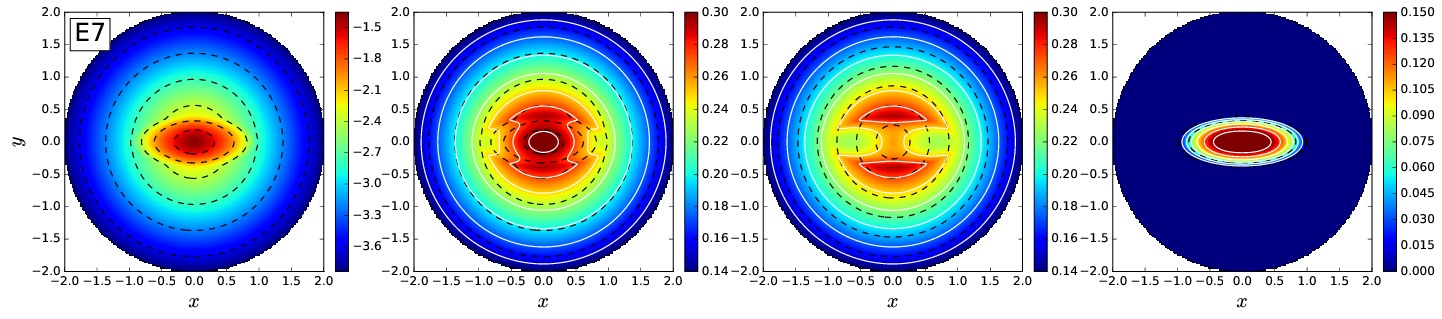}}
       \caption{\label{fig:a3}Models E3, E6, and E7 in \reftab{tab:modelIC}, showing the variation of $\sigma_z$ fields with 
       varying bar thickness $c$ from 0.1 in E3 to 0.5 in E7.} 
\end{figure*}
\begin{figure}[htp]
\centering
       \subfigure{\includegraphics[width=0.45\textwidth]{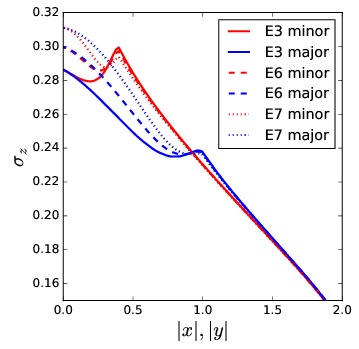}}
       \caption{\label{fig:1Da3}1D $\sigma_z$ profiles along the minor (red) and major (blue) axes of the bars in the models E3 and E6-E7, varying 
            the thickness $c$ of the bar.} 
\end{figure}
\begin{figure}[htp]
\centering
       \subfigure{\includegraphics[width=0.235\textwidth]{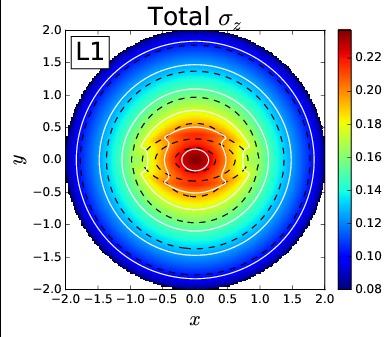}}
       \subfigure{\includegraphics[width=0.235\textwidth]{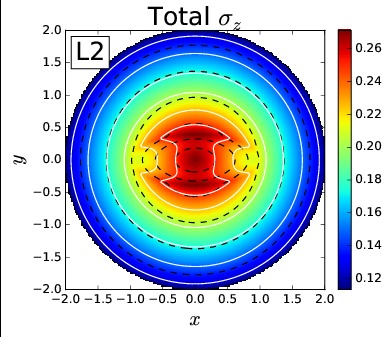}}
       \caption{\label{fig:late}Total $\sigma_z$ maps of the models L1-L2. The disk scale height $h_z$ varies from 
            0.1 in L1 to 0.2 in L2. The overlaid black dashed and white solid curves show the isodensity and $\sigma_z$ contours, respectively.}
\end{figure}

\begin{figure*}[htp]
\centering
       \subfigure{\includegraphics[width=0.95\textwidth]{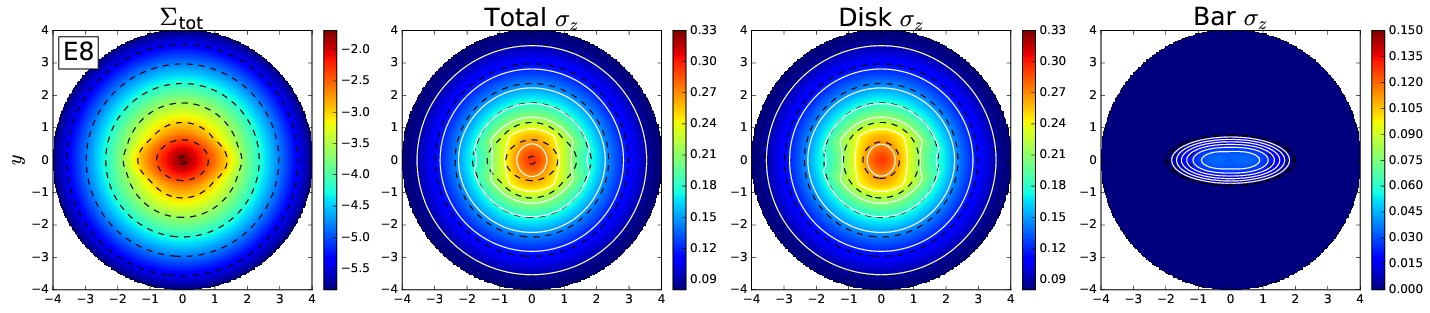}}
       \subfigure{\includegraphics[width=0.95\textwidth]{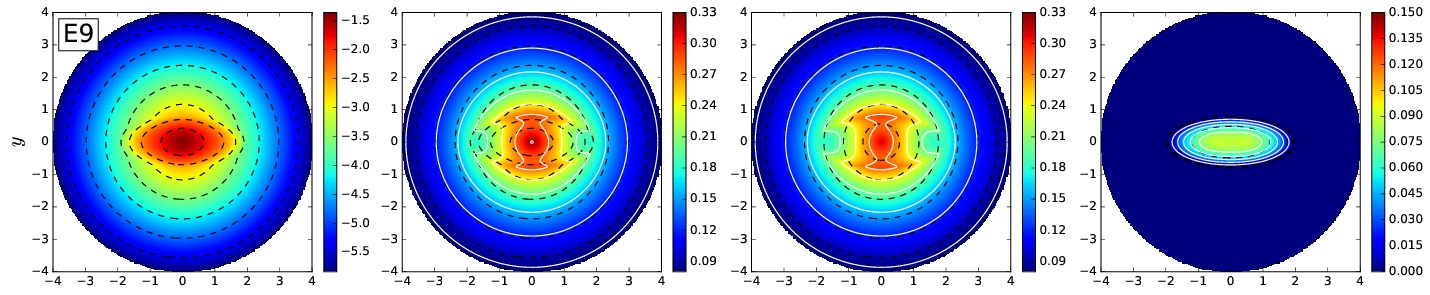}}
       \subfigure{\includegraphics[width=0.95\textwidth]{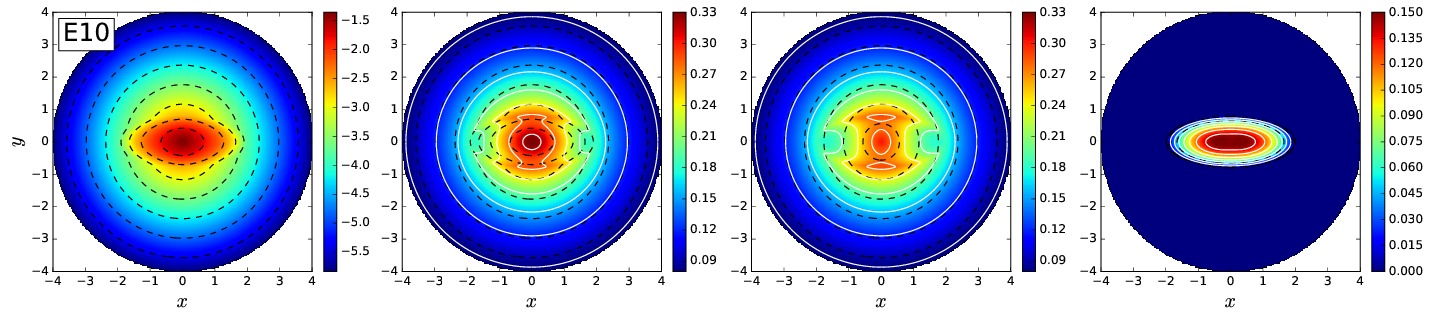}}
\caption{\label{figs:a1}Models E8-E10 in \reftab{tab:modelIC}, showing the $\sigma_z$ maps of the long bar+disk models. The 
            bar is twice the size of the bars in E1-E7, i.e., $a=2.0, b=0.8$. The bar mass in E8-E10 is 
            $0.05M_{\rm D}, 0.2M_{\rm D}$, and $0.2M_{\rm D}$, respectively. The bars in 
            E8 and E9 ($c=0.1$) are vertically thinner than the one in E10 ($c=0.3$).}
\end{figure*}
\begin{figure}[htp]
\centering
      \includegraphics[width=0.45\textwidth]{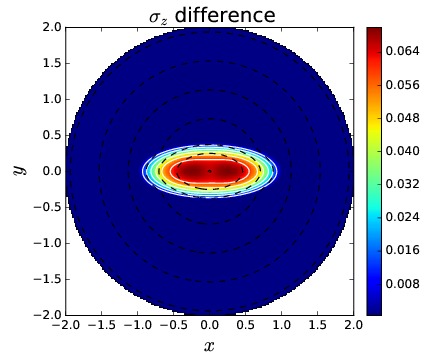}
      \caption{\label{resbar} Map of the $\sigma_z$ difference, obtained by subtracting the total $\sigma_z$ field of E3 from that of the model 
            using a
            vertically exponential density profile with a constant $h_{Bz}=h_z=0.3$. The overlaid black dashed and white solid curves show the 
            isodensity and $\sigma_z$ difference contours, respectively.}
\label{exp3res}
\end{figure}

\subsection{Bar mass}
The pure-disk model, E0, generates an axisymmetric $\sigma_z$ distribution as expected. Adding a bar and keeping its shape  
constant at $a/b/c=1.0/0.4/0.1$, we increase the bar mass by varying $M_\mathrm{B}$ from $0.01M_\mathrm{D}$ in E1 to $0.05M_\mathrm{D}$ in E3.
The total surface density $\Sigma_\mathrm{tot}$ (the first column) and resulting $\sigma_z$ maps of E1-E3 are shown in 
\reffig{fig:BxDs}. The second column shows the total $\sigma_z$ of the bar+disk systems. The disk and the bar $\sigma_z$, 
weighted by the respective surface density, are shown in the third and fourth columns, respectively. As the bar mass increases, 
we can see more prominent \szhuho\ along the minor/major axis of the bar. In order to better appreciate the amplitudes of the \szhuho, we 
plot the 1D $\sigma_z$ profiles along the minor (red) and major (blue) axes of the bars (\reffig{fig:1DBxD}). For comparison, 
the $\sigma_z$ of E0 is overlaid in black. As shown in \reffig{fig:1DBxD}, E0-E3 exhibit almost the same distribution of
$\sigma_z$ at large radii ($R>1.2$), suggesting that the influence of the bar is important only in the inner region, where it dominates. 
With the bar mass increasing, in E2-E3 the $\sigma_z$ values are significantly enhanced on the minor axis of the bar, i.e., $\sigma_z$-humps form, 
which are clearly induced by the nonaxisymmetric bar potential. 
In contrast, $\sigma_z$ values are reduced along the major axis of the bar, thus forming $\sigma_z$-hollows. 
As shown in \reffig{fig:BxDs}, the disk $\sigma_z$ maps (the third column) also show \szhuho\ as in the 
total $\sigma_z$ maps, while none are present in the bar $\sigma_z$ (the fourth column). The oval $\sigma_z$ contours of the bar are 
aligned with the bar. Therefore, surprisingly, although the $\sigma_z$-humps are supported by the bar potential, they are mainly 
present in the disk component, extending beyond the bar along the minor axis. This result is consistent with observations 
\citep{deLorenzoCaceres08, Du16} and simulations \citep{Du16} of S2Bs.

\subsection{Bar ellipticity and thickness} 
In the analysis above, we have shown that even a relatively lightweight bar ($0.05M_\mathrm{D}$) can generate prominent \szhuho. 
However, in most IFU observations and $N$-body simulations of barred galaxies, bars do not usually generate \szhuho. 
Currently, \shuho\ have been seen only in the cases of S2Bs.
In order to identify the condition for generating \szhuho, we study the effect of bar properties ($a, b$, and $c$) on 
$\sigma_z$. As shown in \reftab{tab:modelIC}, fixing $a=1.0$, we vary $b$ and $c$ in models E3-E7. The bar ellipticity 
is varied from 0.8 ($b/a=0.2$) in E4 to 0.4 ($b/a=0.6$) in E5 (\reffig{fig:a2}). The variation of the 1D profiles of
\szhuho\ is shown in \reffig{fig:1Da2}. Here the thickness of the bars is fixed at $c=0.1$, i.e., a 
vertically thin bar. It is clear that a larger ellipticity, i.e., smaller $b/a$, generates more prominent \szhump\ as 
well as hollows.  

In models E3 and E6-E7 we vary the bar thickness from 0.1 in E3 to 0.5 in E7 using a constant $b/a=0.4$ 
(\reffig{fig:a3}). As shown in \reffig{fig:1Da3}, the central $\sigma_z$-drop gradually becomes a 
$\sigma_z$-peak as the bar thickness increases. Because of the enhancement of the bar $\sigma_z$ (the fourth column of
\reffig{fig:a3}), the \szhuho\ become less prominent when the bar is thick. The disk 
$\sigma_z$ is almost unchanged with increasing bar thickness (the third column). As a result, the total $\sigma_z$ contours
of E7 are oval and aligned with the bar. 

In \reffig{fig:late} we show the total $\sigma_z$ maps of the late-type-like models L1-L2. Using a vertically thinner disk in L1 
($h_z=0.1$) and L2 ($h_z=0.2$), $\sigma_z$ is reduced. There are no prominent 
\szhuho\ present in L1. Thus, a vertically even thinner bar is required to generate \szhuho\ in late-type galaxies which are expected 
to be thinner than early-type galaxies.  

\subsection{Bar length}
Using the models E1-E7, we have studied the conditions for generating \szhuho\ in galaxies hosting a typical bar of length $a=1.0$. 
We further examine the $\sigma_z$ features in the long bar+disk models E8-E10 using $a=2.0, b=0.8$ (\reftab{tab:modelIC}). The bar 
in E8 has the same mass ($0.05M_\mathrm{D}$) and thickness ($c=0.1$) as E3; thus, the size increase makes the bar potential 
shallower. As shown in the first row of \reffig{figs:a1}, E8 generates quite round $\sigma_z$ contours as the shallow bar potential 
supports only weak nonaxisymmetric $\sigma_z$ features. 
We set a more massive bar of mass $0.2M_\mathrm{D}$ in E9-E10, varying $c$ from 0.1 in E9 to 0.3 in E10. As shown in \reffig{figs:a1}, 
there are no prominent \szhuho\ present in E9-E10, although the moderately enhanced $\sigma_z$ patterns are somewhat rectangular shaped 
in their outer parts. Such a result suggests that it is more difficult to generate central \szhuho\ in long bars than in short ones. 
Compared to E3, E9 has a similar bar $\sigma_z$ distribution, but its disk $\sigma_z$ is much larger at the center as a result of shallower 
potential. 

\begin{figure*}[htp]
\centering
	\includegraphics[width=0.95\textwidth]{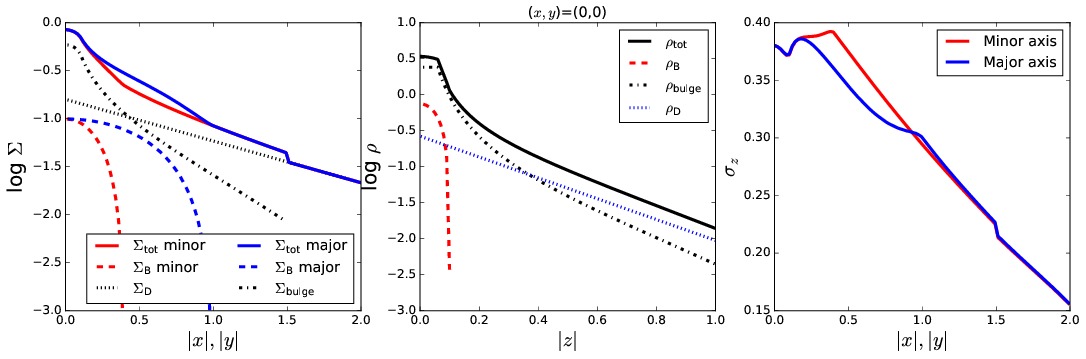}
\caption{\label{fig:bulge} Model E3B, which is identical to the E3 model but with the addition of a bulge of mass 
$M_{\rm bulge}=0.30M_\mathrm{D}$. In the left panel, we plot the logarithmic 
surface density profiles of the total system ($\Sigma_{\rm tot}$), bar ($\Sigma_{\rm B}$), bulge ($\Sigma_{\rm bulge}$), and 
disk ($\Sigma_{\rm D}$), where the red and the blue profiles correspond to the minor and the major axes, respectively. 
The middle panel shows the logarithmic vertical density distributions of the total system 
($\rho_{\rm tot}$), bar ($\rho_{\rm B}$), bulge ($\rho_{\rm bulge}$), and disk ($\rho_{\rm D}$) at the center $(x, y)=(0, 0)$.
The right panel shows the total $\sigma_z$ profiles along the minor (red) and major (blue) axes of the bar.} 
\end{figure*}

\begin{figure*}[htp]
\centering
	\includegraphics[width=0.95\textwidth]{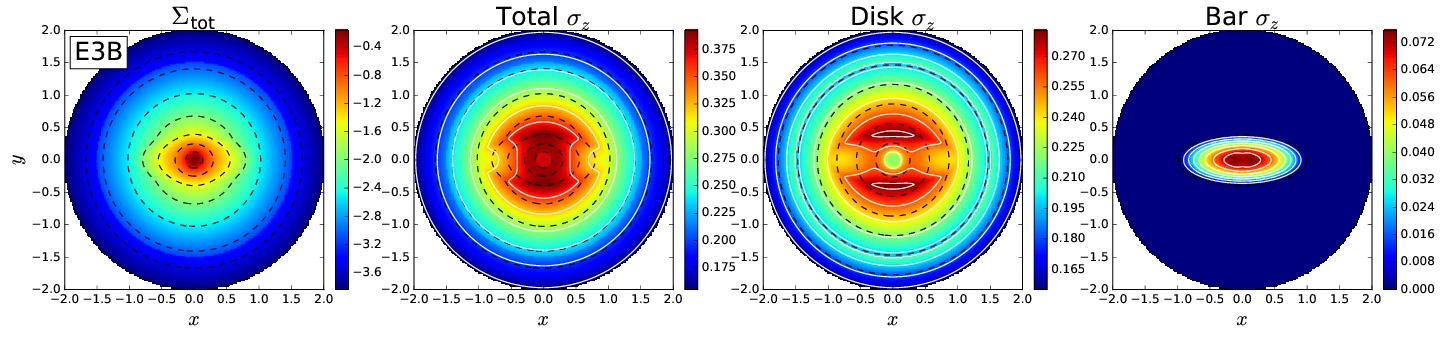}
\caption{\label{figs:bulge}2D $\Sigma_{\rm tot}$ and $\sigma_z$ maps of E3B (\reffig{fig:bulge}).} 
\end{figure*}

\begin{figure}[htp]
\centering
	\includegraphics[width=0.45\textwidth]{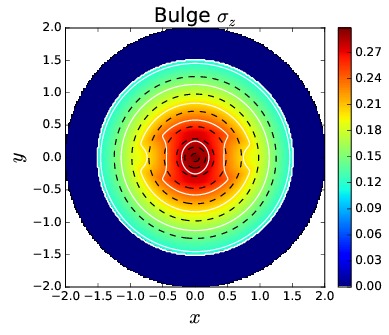}
      \caption{\label{fig:sigzbulge}Bulge $\sigma_z$, weighted by the bulge surface density, i.e., 
            $\sqrt{\Sigma_{\rm bulge}\left<\sigma_z^2\right>_{\rm bulge}/\Sigma_{\rm tot}}$, of the model E3B. 
            The overlaid black dashed and white solid curves show the isodensity and $\sigma_z$ contours, respectively, of the bulge component.}
\end{figure}

\begin{figure}[htp]
\centering
	\includegraphics[width=0.45\textwidth]{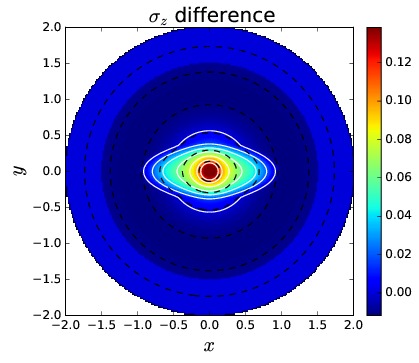}
      \caption{\label{resbulge}Map of the $\sigma_z$ difference, obtained by subtracting the $\sigma_z$ field of the model E3B
            from that of the model using the constant scale height 0.3.
            The overlaid black dashed and white solid curves show the isodensity and $\sigma_z$ difference contours, respectively.}
\end{figure}

\subsection{Vertical density distribution of bars}
\label{vertdens}
According to the analyses above, \szhuho\ are primarily generated by the dynamical response of stars to the 
potential of a vertically thin, sufficiently massive, and relatively short bar.
In photometric observations, we can easily measure the length, ellipticity, and mass of bars, especially in low-inclination 
galaxies, whereas we know little about their vertical density distributions. Although numerical simulations have been widely 
used to study the 3D morphology and orbital structure of bars \citep{Pfenniger84, Martinet&deZeeuw88, Pfenniger&Friedli91, Sellwood&Wilkinson93, Skokos02a, Skokos02b, Patsis02a, Patsis02b, Harsoula&Kalapotharakos09, Valluri16}, 
such theoretical models remain poorly tested by observations. 
We find that the bar $\sigma_z$ is largely determined by the bar thickness, 
while the \szhump\ arising in disks seem to be insensitive to its properties. This suggests that 
$\sigma_z$ can be used as a tracer of the bar thickness. However, our assumption of the vertical density distribution on bars is still 
questionable. The exponential ($\rho(z)\propto {\rm exp}(-z/h_{Bz})$) and isothermal ($\rho(z)\propto {\rm sech^2}(z/2h_{Bz})$) profiles
have also been widely used to approximate the vertical density distribution of real bars for the purpose of estimating the 
bar strength \citep{Buta&Block01, Laurikainen&Salo02, Laurikainen04a, Laurikainen04b, Laurikainen05, Salo15, Diaz-Garcia16}. In this paper 
we considered only models using Ferrers bars. But we have verified that a much thinner bar is also required to generate prominent 
\szhuho\ for a typical bar with vertically exponential or isothermal density profiles. 
In \reffig{resbar} the $\sigma_z$ difference is obtained by subtracting the total $\sigma_z$ of E3 from that of the 
model using a vertically exponential bar whose scale height is the same as the disk, i.e., $h_{Bz}=h_z=0.3$. This helps to quantify 
how the variation of vertical density distribution of bars affects $\sigma_z$. 
As the difference in the total $\sigma_z$ is almost only caused by the difference in the bar $\sigma_z$, 
the positive $\sigma_z$ difference closely traces the thin bar in E3. The $\sigma_z$ difference outside of the bar is close to zero. 
Thus, the $\sigma_z$ difference may be used as a diagnostic of the relative thickness of bars and their host disks.

\section{The effect of bulges}
\label{bulge}

In early-type disk galaxies, a large fraction of the luminosity comes from a massive spheroidal bulge. Having large random 
motions, bulges may affect the $\sigma_z$ features significantly where \szhuho\ arise. In order to study the effect of a bulge on 
$\sigma_z$, we add an oblate, spheroidal power-law bulge in E3 using Equation (2.207) of \citet{Binney&Tremaine08},
\begin{equation}
\rho_{\rm bulge}(R, z) = \rho_{\rm b0} m^{-\alpha_b} e^{-m^2/r_b^2} \         \   (R \le r_b),
\end{equation}
where $m=\sqrt{R^2 + z^2/q_b^2}\ge 0.1$. We set $\alpha_b=1.8$ and $q_b=0.6$ \citep[][Section 10.2.1]{Binney&Merrifield98}. In order to 
avoid the singularity at the center, $\rho_{\rm bulge}$ is set to be constant at $m\le0.1$. The bulge is truncated at $r_b=1.5$, so the bar 
is fully embedded in the bulge. The bulge mass is set to $0.3M_{\rm D}$. This bar+disk+bulge system is named as E3B.
Along the minor and major axes of the bar, the profiles of the surface density and $\sigma_z$ are shown in 
the left and the right panels, respectively, of \reffig{fig:bulge}. The vertical density profiles at the center are 
shown in the middle panel. It can clearly be seen that the bulge is more massive than the bar at any position.   

The 2D $\Sigma_{\rm tot}$ and $\sigma_z$ maps are shown in \reffig{figs:bulge}. As shown in the total $\sigma_z$ map, the presence 
of the bulge significantly raises $\sigma_z$ in the bulge region. It generates similar \szhump\ along the minor axis to the 
simulations in \citet{Du16}, where the central $\sigma_z$-drop becomes more flat-topped (the rightmost panel of \reffig{fig:bulge}) than E3.
In the bulge (\reffig{fig:sigzbulge}) the central $\sigma_z$ contours are slightly oval and perpendicular to the bar. Such a result suggests 
that in the bar potential the bulge component is not as responsive as the disk component, thus generating weaker nonaxisymmetric 
$\sigma_z$ features. The main influence of the bulge is to make the \szhump\ less obvious by enhancing the central $\sigma_z$, 
thus hiding the \szhuho\ to a certain degree. By varying the thickness of the bar in such bar+disk+bulge systems (not shown here), we verify again
that the bar needs to be much thinner than its host disk in order to generate visible \szhuho.

As presented in \refsec{vertdens}, the positive $\sigma_z$ difference can be used as a tracer of the thin bar. 
As shown in \reffig{resbulge}, we obtain the $\sigma_z$ difference of E3B using the same 
approach. We firstly regenerate the vertical density distribution of the whole system with the exponential function using a constant 
scale height 0.3, which is used to recalculate $\sigma_z$. Then the original $\sigma_z$ field is subtracted from the recalculated $\sigma_z$. 
Outside the bulge the $\sigma_z$ difference is close to zero; the regions having positive and negative $\sigma_z$ difference trace well 
the intrinsic face-on morphology of the thin bar and thick bulge, respectively. 
\citet{Debattista05} showed that the fourth-order Gauss-Hermite moment $h_4$ can be used as a kinematic diagnostic for bulges in nearly 
face-on galaxies. Here we show that $\sigma_z$ difference can be used as an alternative kinematic diagnostic of the stellar 
components having different thickness, e.g. thin bars and thick bulges, in barred galaxies. 

\section{$N$-body simulations of nuclear and large-scale bars}
\label{simulation}

\begin{figure*}[htp]
\centering
        \subfigure{\includegraphics[width=0.98\textwidth]{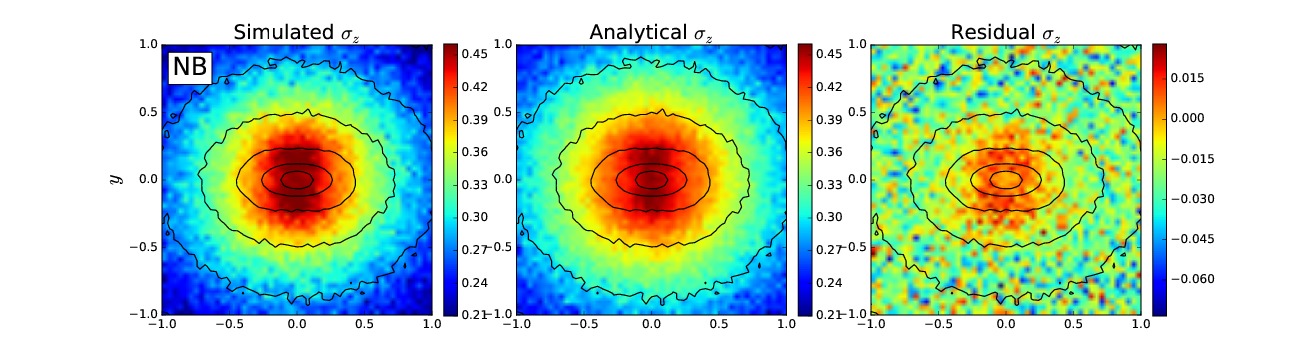}}
        \subfigure{\includegraphics[width=0.98\textwidth]{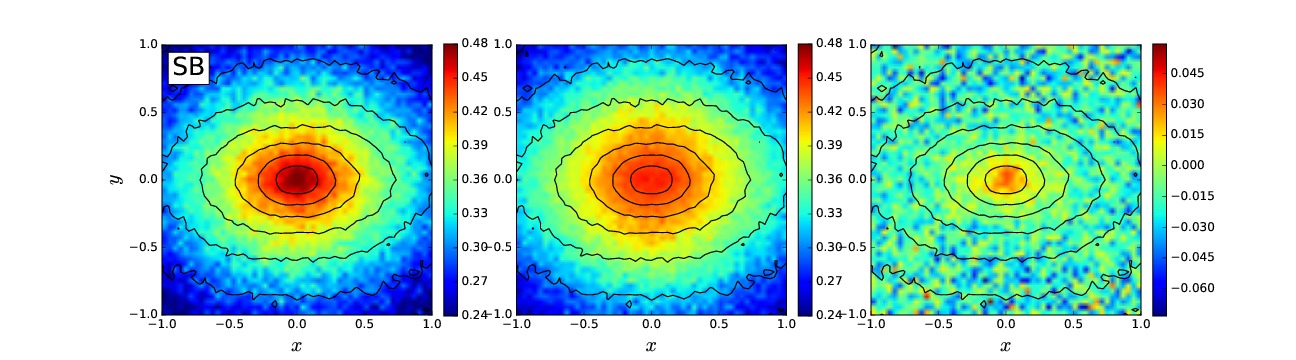}}
        \caption{\label{fig:NBSB}Maps of the simulated $\sigma_z$ (left) and the analytical $\sigma_z$ (middle) calculated using \refeq{eq:zint}, 
            and the residual 
            $\sigma_z$ (right) of the nuclear-barred (NB, top) and the large-scale single-barred (SB, bottom) models.  
            The residual $\sigma_z$, obtained by subtracting the analytical 
            $\sigma_z$ from the simulated $\sigma_z$, corresponds to the contribution of the anisotropic pressure in the simulations. 
            We fix the color bars of the left and middle columns. A much smaller range is used in the right column, as the maximum 
            residual $\sigma_z$ is at the $\sim 5-10\%$ level of the simulated $\sigma_z$.
            The surface density contours are overlaid in black, separated in equal intervals in logarithmic space.} 
\end{figure*}
\begin{figure*}[htp]
\centering
\includegraphics[width=0.98\textwidth]{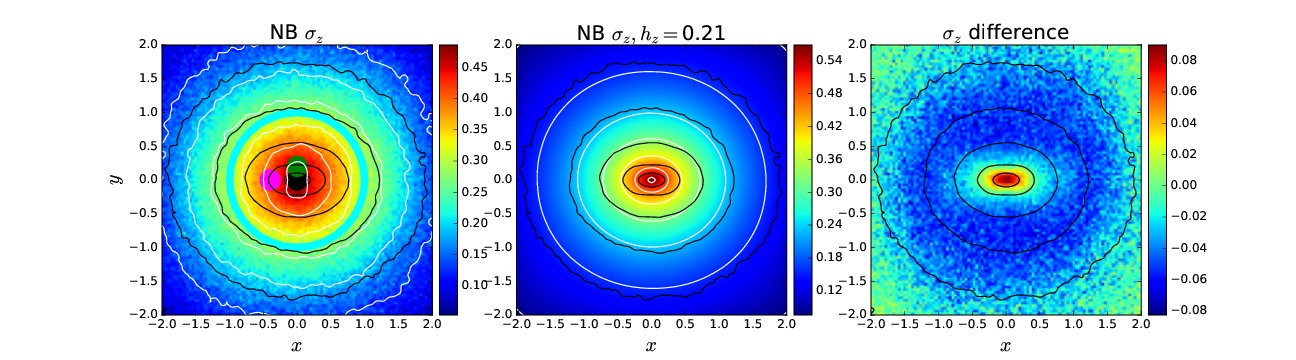}
\includegraphics[width=0.98\textwidth]{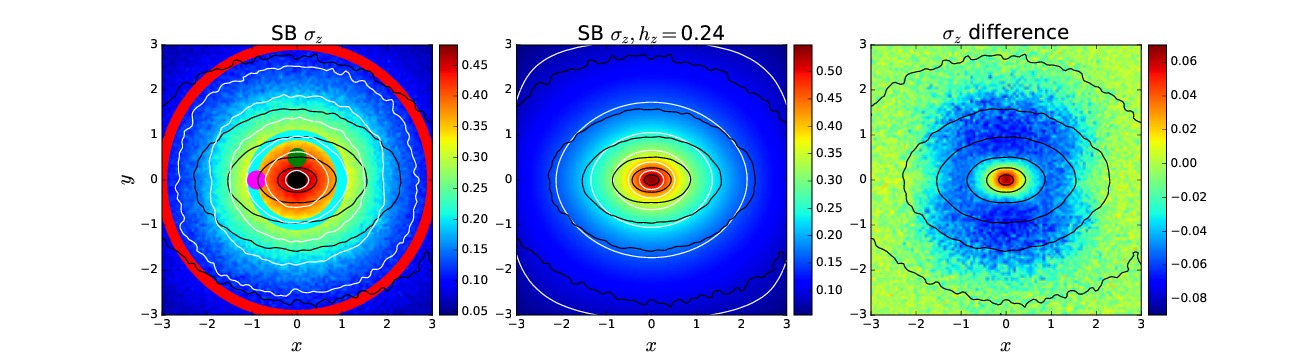}
\caption{\label{fig:sim} Numerically calculated $\sigma_z$ using different vertical density distributions and their $\sigma_z$ difference 
      (right) maps. Based on \refeq{eq:zint}, the $\sigma_z$ maps in the left panels are calculated using the original 3D density distributions 
      from the 
      NB (top) and SB (bottom) simulations. In the middle panels, without changing the surface density, we recalculate the $\sigma_z$ maps by 
      using a vertically exponential profile with a constant scale height. The constant scale heights are set to the linear fitting 
      of the vertical density distributions of the NB ($h_z=0.21$) and SB ($h_z=0.24$) outer disks (\reffig{fig:fithz}), respectively. 
      The right panels show the $\sigma_z$ difference between the $\sigma_z$ calculated using the original density and that using a 
      vertically exponential density distribution. The smoothed surface density 
      and $\sigma_z$ contours are overlaid in black and white, respectively. In the left panels, the colored dots and annuli mark the 
      regions we use to average the vertical density profiles in \reffig{fig:fithz}. In the NB model the red annulus at $R=3.0$ is 
      beyond the boundary of the image in the NB model, thus not shown here.}
\end{figure*}

\begin{figure*}[htp]
\centering
\includegraphics[width=0.45\textwidth]{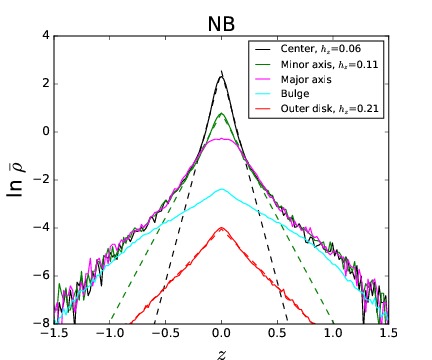}
\includegraphics[width=0.45\textwidth]{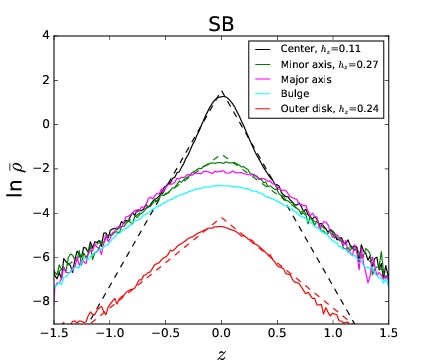}
\caption{\label{fig:fithz}Average density (ln$\bar{\rho}$) profiles of the NB (left) and the SB (right) models along 
            the $z$ direction. The profiles 
            correspond to the vertical density distributions of the center (black), minor (green), and major (magenta) axes of the bar, 
            B/P bulge (cyan), and outer disk (red) regions. The average regions are marked with the filled dots and annuli in the same color as in 
            \reffig{fig:sim}. The dashed profiles represent the extrapolated linear fitting of the ln$\bar{\rho}$ profiles at the 
            outer disk (red) and the center (black) and minor axis (green) of the bar. The fitted scale heights are given in the legend.} 
\end{figure*}

Using the bar+disk(+bulge) models, we have demonstrated that a vertically thin bar is required to generate \szhuho\ in 
barred galaxies. Such models allow us to study the effect of any single parameter by fixing the others. 
However, it is difficult to measure the 3D density distribution, especially perpendicular to the disk plane, 
in real galaxies. In order to verify the analytical results above, we study the 3D density distribution of self-consistent $N$-body 
simulations from \citet{Du15, Du16}. The unique advantage of simulations is that the 3D density distribution is completely known. 

As shown in \reffig{fig:NBSB}, we have studied two representative cases, namely, a nuclear-barred simulation ``NB'' (the top row) and a 
large-scale single-barred simulation ``SB'' (the bottom row). Here we briefly summarize the properties of the NB and the SB models 
\citep[see more details in][]{Du15}. Starting from a pure exponential disk with 4 million particles, the models were evolved 
using a 3D cylindrical polar grid code, \galaxy\ \citep{Sellwood&Valluri97, Sellwood14}. The unit system of the simulations is the same as the 
analytical models in \refsec{results}. By reducing the Toomre-$Q$ in the inner region, the initial inner disk generally 
triggers a significant nuclear bar instability, forming a nuclear-barred galaxy or double-barred galaxy \citep[Fig. 1 in][]{Du15}. 
Possibly because of the heating of spirals driven by the nuclear bar, the outer disk in the NB model becomes too hot to form a bar
\citep{Athanassoula&Sellwood86, Du15}. 
After reaching a quasi-steady state, the semimajor axis of the nuclear bar extends to $\sim0.7$ of the initial $h_R$, making it a quite short bar. 
The NB model exhibits similar \shuho\ to the standard S2B \citep{Du15, Du16}. Thus, the outer bar is not a necessary condition for generating 
\shuho. It is worth emphasizing that the initial thickness of the NB model is smoothly lowered to 0.05 inside $R<1.0$ from 0.1 in the 
outer region. In this case the NB model generates more prominent \shuho\ (the top left panel of 
\reffig{fig:NBSB}) than the cases of using a radially constant thickness of 0.1.

Using a dynamically hotter initial inner disk normally leads to a large-scale single bar. The SB model here is exactly 
the same model as in Fig. 6 of \citet{Du16} where the $\sigma_z$ contours are oval and aligned with the bar
(the bottom left panel of \reffig{fig:NBSB}). The semimajor axis of the bar in the SB model is $\sim 3.0$.
Both the NB and the SB models are thickened in their inner regions ($R\sim 1.5$), where boxy/peanut (B/P) bulges possibly form as 
seen from the edge-on view. Thus, in the NB model the nuclear bar is embedded in the host bulge. 

\subsection{Quantifying the uncertainty due to the anisotropic pressure in barred galaxies}
\label{aniso}

In this study we have assumed that the velocity cross-terms are unimportant in the vertical dynamics (see \refsec{Jeans}), which is 
generally considered to be a good approximation in axisymmetric systems. 
However, this assumption is not obviously justified for nonaxisymmetric bars that induce large streaming motions, 
possibly causing a systematic error in the $\sigma_z$ calculation. In order to quantify the anisotropic pressure caused by the 
velocity cross-terms, we apply the vertical kinematic estimation to the NB and SB models. The density distribution and associated vertical 
force from the simulations are used to calculate 
the analytical $\sigma_z$ (middle column of \reffig{fig:NBSB}). By subtracting the analytical $\sigma_z$ from the simulation's actual
$\sigma_z$ (left column), we obtain the residual $\sigma_z$ (right column) that corresponds to the contribution of the 
anisotropic pressure. We have verified that the contribution of the cross-terms (the second term on the right-hand side of 
\refeq{eq:int}) computed directly is almost the same as that of the residual $\sigma_z$ here. 

The residual $\sigma_z$ is roughly equal to zero all over the disk for the SB model. Only in the very central 
region is the residual $\sigma_z$ positive at the $\sim10\%$ level, which has no effect on $\szhuho$. In the NB model there is an 
extensive positive residual $\sigma_z$ ($\sim5\%$ level) along the minor axis of the bar, which is possibly related to the 
elongated streaming motions in the nuclear bar. We have checked the standard S2B model 
as well, in which the anisotropic pressure enhances $\sigma_z$ values along the minor axis of the inner bar at a similarly low 
level to the NB model. The maximum ellipticity of the simulated bars here reaches $\sim0.6$. In observations some late-type 
bars can be very narrow and strong, in which case the importance of the anisotropic pressure may increase. Therefore, a cautious 
conclusion is that the anisotropic pressure is negligible in galaxies containing a normal or weak bar.

\subsection{A kinematic diagnostic of vertical thickness: $\sigma_z$}

As the anisotropic pressure is negligible, the $\sigma_z$ features of the NB and SB models are mainly determined by their 3D density 
distributions and associated potentials. In this section, we investigate whether a vertically thin bar exists in the NB model, which  
our analysis in \refsec{results} suggests is a necessary condition for generating \szhuho\ (\reffig{fig:sim} and \ref{fig:fithz}). The SB model is shown for comparison purposes.

In order to reduce the noise, we select the colored regions (annuli and filled dots in the left column of \reffig{fig:sim}) to average the vertical 
density distribution. In both the NB (top) and SB (bottom) models the filled dots correspond to 
the minor axis (green), major axis (magenta), and center (black) of their bars. The cyan (at $R=1.0$) and the red (at $R=3.0$) annuli represent the 
B/P bulge and the outer disk regions, respectively. The red outer disk annulus in the NB model is not 
shown, as the computed region at $R=3.0$ is beyond the boundary $[-2.0, 2.0]$ of the image. In \reffig{fig:fithz} the average density distributions
of each of these regions are indicated by the solid profiles using the same color. The dashed profiles represent the extrapolated linear fitting 
of the ln$\bar{\rho}$ profiles at each region. 
In the bar regions we only fit the region close to the 
midplane where the bar should dominate (for the NB $z\in [-0.3,0.3]$, while for the SB $z\in[-0.5, 0.5]$). The density profiles on the 
major axis of the bar (magenta) and the B/P bulge (cyan) cannot be fitted by a linear relation. The fitted scale height values at each region 
are given in the legend. The nuclear bar of the NB model ($h_z=0.11$ at the minor-axis area) is vertically thinner than the host disk 
($h_z=0.21$), which agrees well with the analytical expectation for generating \szhuho. In contrast, the bar of the SB model is as 
thick as the disk, except for its very central region; thus, it exhibits no \szhuho.

We calculate the $\sigma_z$ difference using the same approach as in Sections \ref{vertdens} and \ref{bulge}. In the left panels of 
\reffig{fig:sim} we show the numerically calculated $\sigma_z$ using the original 3D density distributions of the NB and SB models.
In the middle panels $\sigma_z$ is recalculated using the vertically exponential profiles with a constant scale height. 
The scale heights used here are set to the linear fitting results of the outer disks (NB $h_z=0.21$; SB $h_z=0.24$). 
Then the $\sigma_z$ difference maps (right panels) are obtained by subtracting the $\sigma_z$ maps in the left panels 
from those in the middle panels. In this case, the nonzero $\sigma_z$ difference represents the difference of vertical thickness from 
the outer disk. The $\sigma_z$ difference is roughly equal to zero in the outer disk. In the NB model the $\sigma_z$ difference 
is qualitatively consistent with the bar+disk+bulge model (\reffig{resbulge}). In the thin bar region the $\sigma_z$ difference 
is positive at the $\sim10-20\%$ level, while in the thick B/P bulge region it turns out to be 
negative ($\sim15-30\%$ level). As visually confirmed from the edge-on view, the NB model hosts a nearly boxy bulge of radius $R\sim1.5$. 
The negative $\sigma_z$ difference traces the face-on morphology of such a boxy bulge. In the bottom right panel, the B/P bulge of the 
SB model should correspond to the peanut-shaped negative region ($\sim25\%$ level) in the $\sigma_z$ difference map. A positive 
$\sigma_z$ difference only appears at the very central region (marked with the filled black dot in \reffig{fig:sim}) where $\bar{\rho}$ is 
peaked around the midplane. 

In conclusion, the $\sigma_z$ difference seems to be a good kinematic diagnostic for the stellar components having different thickness, e.g. 
thick bulges and thin bars. It may shed new light on the 3D geometry of bars and bulges in the face-on views of barred galaxies. 
It is worth emphasizing that, for real galaxies, $h_z$ is generally estimated from either the empirical relation in \citet{deGrijs98} or 
the observed $\sigma_z$ in the outer disk by assuming a reasonable mass-to-light ratio. In practice, the estimation of $h_z$ still has 
a large uncertainty, and the mass-to-light ratio is not constant. This may cause large errors 
in the estimation of surface density. The practicality of this method will be tested in future work.

\section{Summary}
\label{summary}

By applying the vertical Jeans equation to a group of well-designed bar+disk(+bulge) models, we have systematically investigated the 
$\sigma_z$ properties of barred galaxies from a purely dynamical point of view. The main conclusions can be summarized as follows:
\begin{enumerate}
\renewcommand{\labelenumi}{(\theenumi)}
\item Bars can dynamically induce significant nonaxisymmetric $\sigma_z$ features, either \szhuho\ or oval $\sigma_z$ contours aligned with bars. 
The properties of $\sigma_z$ features are tightly related with the properties of bars, i.e., mass, length, ellipticity, and thickness. 
Generally, thick or long bars are more likely to generate oval $\sigma_z$ contours aligned with bars. 
\item We found that vertically thin bars can not only reduce $\sigma_z$ along the major axis of bars but also enhance $\sigma_z$ along the minor 
axis, thus generating \szhuho. Such \szhuho\ can explain the \shuho\ appearing in the kinematic observations of double-barred galaxies. 
\item As a dynamical response of stars to the potential of bars, the amplitude of $\sigma_z$-humps is proportional to 
the mass and ellipticity of bars, while it is almost independent of the bar thickness. $\sigma_z$-humps are mainly present in host
disks, thus extending beyond bars. A thin bar mainly reduces $\sigma_z$ in the bar region, thus generating $\sigma_z$-hollows.
\item We showed that \szhuho\ are preferentially found in galaxies harboring a short bar, e.g. inner bars of double-barred galaxies and 
single nuclear bars. \shuho\ have been commonly observed in double-barred galaxies, while their frequency in 
nuclear-barred galaxies is still unclear. In long bar cases \szhuho\ are less frequent, possibly because volume expansion makes bar 
potential shallower. 
\item Using the bar+disk+bulge models, we show that the primary effect of a thick bulge is to make the $\sigma_z$-humps weaker 
by enhancing the central $\sigma_z$. Thus, \shuho\ should not be explained by the contrast of dynamically cold bars and hot bulges as 
proposed in previous analysis.
\end{enumerate}

In IFU observations, an increasing number of \shuho\ features have been identified in nearby S2Bs 
\citep{deLorenzoCaceres08, deLorenzoCaceres13, Du16}. \citet{Du16} presented self-consistent S2B simulations that match 
the kinematic observations of S2Bs. In this paper, we demonstrate that the existence of a vertically thin bar in the nuclear-barred simulation 
(NB) generates such \shuho\ in small-scale (nuclear) bars. 
The interaction of multiple bars should play a minor role in generating \shuho. 
The ubiquitous presence of \shuho\ in S2Bs indicates that inner bars are vertically thin structures. Thus, it
suggests that inner bars either are not prone to thickening or they are younger structures formed in dynamically cold nuclear disks. 
However, the detailed stellar population analysis of S2Bs showed that inner bars are not young structures 
\citep{deLorenzoCaceres12, deLorenzoCaceres13}. In our simulations vertically thin bars also last for more than 5Gyr. Thus, we 
propose that inner bars are weakly thickened after forming in initial nuclear disks.


As embedded in galactic central regions, the vertical thickness of bars is rarely measured in real galaxies. In low-inclination cases, it 
is also very difficult to identify the morphology of bulges. An implication of this work is that $\sigma_z$ may trace the stellar 
components having different thickness, e.g. thin bars and thick bulges. It may provide a novel perspective on the 3D geometry of bars and bulges 
from IFU surveys for nearly face-on galaxies. 

\begin{acknowledgments}
M.D. thanks the Jeremiah Horrocks Institute of the University of
Central Lancashire for their hospitality during a 3-month visit
while this paper was in progress. 
The research presented here is partially supported by the 973 Program of
China under grant no. 2014CB845700, by the National Natural Science
Foundation of China under grant nos. 11333003 and 11322326, and by the
Strategic Priority Research Program ``The Emergence of Cosmological
Structures'' (no. XDB09000000) of the Chinese Academy of Sciences. We
acknowledge support from a Newton Advanced Fellowship no. NA150272 awarded by
the Royal Society and the Newton Fund. This work made use of the facilities
of the Center for High Performance Computing at Shanghai Astronomical
Observatory. V.P.D. is supported by STFC Consolidated grant no. ST/J001341/1.
V.P.D. was also partially supported by the Chinese Academy of Sciences
President's International Fellowship Initiative Grant (no. 2015VMB004).
A.d.L.-C. acknowledges support from the CONACyT-125180, DGAPA-IA100815 and DGAPA-IA101217 projects.
\end{acknowledgments}

\bibliographystyle{apj}
\bibliography{BarModel}

\end{document}